\documentclass[prb,
superscriptaddress,showpacs,amsmath,amssymb]{revtex4}

\begin{document}

\title{Tsallis-Cirto entropy of black hole and black hole atom}

\author{G.E.~Volovik}
%\affiliation{Low Temperature Laboratory, Aalto University,  P.O. Box 15100, FI-00076 Aalto, Finland}
\affiliation{Landau Institute for Theoretical Physics, acad. Semyonov av., 1a, 142432,
Chernogolovka, Russia}

\date{\today}

\begin{abstract}
The quantum tunneling processes related to the black hole determine the black hole thermodynamics. The Hawking temperature is determined by the quantum tunneling processes of radiation of particles from the black hole. On the other hand, the Bekenstein-Hawking entropy of the black hole is obtained by consideration of the macroscopic quantum tunneling processes of splitting of black hole to the smaller black holes. These tunneling processes also determine the composition rule for the black hole entropy, which  coincides with the composition rule for the non-extensive Tsallis-Cirto $\delta=2$ entropy. This composition rule suggests that the mass spectrum of the black hole is equidistant, $M=NM_0$. Here $N$ is an integer number and $M_0=\sqrt{2}m_{\rm P}$ is the mass quantum expressed via the reduced Planck mass $m_{\rm P}$. The  Bekenstein–Hawking entropy of the black hole with mass $M=NM_0$ is $S_{\rm BH}(N)=N^2$. 
\end{abstract}
\pacs{
}

\maketitle

  \tableofcontents

\section{Introduction}

The Bekenstein-Hawking area law for the black hole entropy is non-extensive, and there are many attempts 
to find possible alternatives to the Bekenstein-Hawking entropy using the Renyi and Tsallis statistics and the other types of generalized entropy, see e.g. Refs. \cite{Odintsov2021,Odintsov2022,Anand2024} and references therein.
In particular, it was suggested that the non-additivity of the Bekenstein–Hawking entropy of the Schwarszchild black holes, $S_{\rm BH}(M)=4\pi G M^2$, can be treated in terms of the non-extensive Tsallis-Cirto entropy.\cite{TsallisCirto2013,Tsallis2020,Dabrowski2024} However, according to Tsallis,\cite{Tsallis2020} some nontrivial physical points still remain open. Also the application of the generalized entropy leads to problems with the Hawking temperature \cite{Odintsov2022}.

Here we consider the black hole configurational space, where only the processes between the black holes of different masses are taken into account, i.e. without any matter components. We show that the rare quantum tunneling processes of the splitting of the black hole into the smaller black holes with the conservation of the total mass provide the non-extensive composition law for the black hole entropy. This composition rule coincides with that for the Tsallis-Cirto $\delta$-entropy with $\delta=2$. 
It appears that the Tsallis-Cirto  $\delta =2$ entropy is probably the only example of the generalized entropy, which exactly coincides with the Bekenstein-Hawking entropy and is consistent with the black hole thermodynamics.

This also suggests the specific quantization of the black hole states, which is different from the Bekenstein quantization of the black hole entropy. Now it is the mass spectrum of the black hole, which is equidistant, $M=M_0 N$, where $N$ is an integer number and $M_0=\sqrt{2}m_{\rm P}$ is the mass quantum expressed via the reduced Planck mass, $m_{\rm P}=1/\sqrt{8\pi G}$. This actually means that the black hole with mass $M$ can be considered as an ensemble of $N$ micro black holes, which play the role of the black hole "atoms". The  Bekenstein–Hawking entropy of the black hole with $N$ "atoms" is $S_{\rm BH}=N^2$. 

The plan of the paper is the following. 

In Sec. \ref{HawkingRadiation} we recall the Hawking radiation as the process of quantum tunneling of particles from the black hole. The important element is the back reaction due to decrease of the black hole mass after emission. 

In Sec. \ref{TunnelingEmission} the tunneling process is extended to the macroscopic quantum tunneling, which describes the processes of splitting of black holes into the smaller parts. When the back reaction is properly included, these processes are fully determined by the Bekenstein–Hawking entropies of the participating black holes. This suggests that the black hole thermodynamics can be described in terms of the ensemble of black holes. This also demostrates that the Hawking temperature and the Bekenstein-Hawking entropy can be derived by consideration of the macroscopic quantum tunneling between the black hole states. 

In Sec. \ref{BlackHoleEnsemble} it is shown that the rare fluctuations -- the exponentially suppressed processes of transformation of the black holes by macroscopic quantum tunneling -- are the main reason of the non-additivity of the Bekenstein–Hawking entropy. These rare processes play the role of intermittency in the chaotic processes, which breaks the additivity of entropy. If black hole splitting were allowed at the classical level, i.e. at $\hbar\rightarrow 0$, the black hole entropy would be extensive. However, at the classical level such a process is forbidden.

In Sec. \ref{TCentropy} it is shown that the non-additivity of the Bekenstein–Hawking entropy is equivalent to the non-additivity of the Tsallis-Cirto $\delta$-entropy with $\delta=2$. 

In Sec. \ref{QuantumMassSec} it is shown that the black hole can be considered as the ensemble of "atoms  of the black hole matter". Here the role of an atom is played by the micro black hole with the Planck-scale mass $M_0=1/\sqrt{4\pi G}$. This corresponds to quantization of the black hole mass in terms of the mass quantum, $M=NM_0$. The micro black holes with the Planck scale masses have been discussed by Hawking,\cite{Hawking1971} see also Refs. \cite{Chen2005,Faraoni2017,Rasanen2019,Profumo2024,Rovelli2024,Haug2024,Ong2024} and Markov's maximons.\cite{Markov1967}

In Sec. \ref{RNsection} it is shown that this approach is also valid for Reissner-Nordstr\"om (RN) black holes with two horizons. This is because the entropy of Reissner-Nordstr\"om black hole depends only on its mass, and thus the charged black hole has the same quantization rules for mass and entropy as the Schwarszchild black hole, $M=NM_0$ and $S_{\rm RN}(N)=N^2$. it is also suggested that in addition to the elementary black holes, one may have the white-hole "atoms" with the same value $M_0$ of the elementary mass. Then the entropy of the white hole with mass $M=NM_0$ is $S_{\rm WH}(N)=-N^2$.

%In Conclusion section \ref{ConclusionSec} 

\section{Hawking radiation as quantum tunneling}
\label{HawkingRadiation}

Let us start with semiclassical tunneling description of the Hawking radiation of particles from the black holes.\cite{Wilczek2000,Srinivasan1999,Volovik1999}
The rate of emission of the particle with energy $\omega$ from the black hole with mass $M$ has the following exponential law under condition $T_{\rm H}\ll \omega \ll M$:
\begin{equation}
w(\omega, M)\propto \exp{\left(-\frac{\omega}{T_{\rm H}}\right)} \,\,,\,\, T_{\rm H}=\frac{1}{8\pi GM}
\,.
\label{tunneling}
\end{equation}
Thus the quantum tunneling process provides the Hawking temperature $T_{\rm H}$, which determines is the rate of emission.

The further important step was made by Parikh and Wilczek,\cite{Wilczek2000} who obtained the correction to the Hawking radiation. This correction is caused by the back reaction --  the reduction of the black hole mass after emission:
 \begin{equation}
w(\omega, M-\omega)\propto \exp{\left(-8\pi G\omega\left(M-\frac{\omega}{2}\right)\right)}
\,.
\label{tunnelingMomega}
\end{equation}

This back reaction can be related to the decrease of the Bekenstein-Hawking entropy $S_{\rm BH}(M)=4\pi GM^2$ of the black hole after emission.\cite{Kraus1997} One can consider the Hawking radiation as the rare effect caused by thermodynamic fluctuations, while the latter can be described in terms of entropy change.\cite{Landau_Lifshitz} 
Then the rate of Hawking radiation in Eq.(\ref{tunnelingMomega}) can be described in terms of the difference of the  black hole entropies before and after emission of a particle:
\begin{equation}
w(\omega, M-\omega)\propto \exp{\left[S_{\rm BH}(M-\omega)-S_{\rm BH}(M)\right]}
\,.
\label{EntropyDifference}
\end{equation}
This shows that the quantum tunneling process may serve as the source of both the Hawking temperature and the Bekenstein-Hawking entropy. This is supported by consideration of the macroscopic quantum tunneling which describes the emission of black holes.\cite{Volovik2022}

\section{Emission of black holes as macroscopic quantum tunneling}
\label{TunnelingEmission}

The process of emission a small black hole, which can be  considered as a kind of particle, contains the similar element of back reaction.\cite{Volovik2022} However, the emitted black hole "particle" has the nonzero entropy. As a result, as distinct from the radiation of the point particle, the rate of emission of a small black hole is enhanced by the entropy of the emitted black hole.\cite{HawkingHorowitz1995} If we consider the emission of the black hole of mass $m$,
 then the probability of emission contains the extra term compared to the emission of particles -- the entropy of the emitted black hole, $S_{\rm BH}(m)=4\pi G m^2$:
 \begin{eqnarray}
w(m,M-m)\propto \exp{\left[-8\pi Gm\left(M-\frac{m}{2}\right)+ 4\pi G m^2\right]} =\exp{\left[-8\pi Gm(M-m)\right]}\,.
\label{HoleEmession1}
\end{eqnarray}
The Eq.(\ref{HoleEmession1}) demonstrates  that in general the rate of the splitting of the black hole into two black holes obeys the following rule:
 \begin{eqnarray}
w(M\rightarrow M_1+M_2)\propto \exp{\left[S_{\rm BH}(M_1)+S_{\rm BH}(M_2)-S_{\rm BH}(M_1+M_2)\right]}
\,.
\label{HoleEmession2}
\end{eqnarray}
This shows that the processes which involve the black holes in the absence of matter, such as splitting and merging of the black holes, are determined by the entropies of black holes. These processes form the configurational space of the black holes, which is the source of the generalized entropy of black holes.

\section{Configurational space in the black hole ensemble}
\label{BlackHoleEnsemble}

Eq.(\ref{HoleEmession2}) demonstrates the reason, why the black hole entropy does not satisfy the additivity condition,  $S(A,B)=S(A) +S(B)$, which is valid for the thermodynamics of the extensive systems. In the systems, which obey the extensive thermodynamics, the entropy is proportional to the volume of the system. The splitting of the system with volume $V$ in two parts with volumes $V_1+V_2=V$ does not change the total entropy of the system,  $S(V_1+V_2)=S(V_1) +S(V_2)$.

On the contrary, according to Eq.(\ref{HoleEmession2}), the splitting of the black hole into two smaller black hole is the consequence of the rare process of the macroscopic quantum tunneling transition. The calculations of the tunneling exponent demonstrate that this rare process can be considered as the thermodynamic fluctuation, which rate according to Landau and Lifshitz\cite{Landau_Lifshitz} is determined by the entropy change of the system, $w\propto e^{-\Delta S}$ with
\begin{eqnarray}
\Delta S=S_{\rm BH}(M_1+M_2)-S_{\rm BH}(M_1)-S_{\rm BH}(M_2)>0
\,.
\label{nonadditive}
\end{eqnarray}
Considering the rates of different processes of splitting of black holes, one comes to the following non-additive composition rule for the black hole entropies:
\begin{equation}
S_{\rm BH}(M_1 +M_2)= \left( \sqrt{S_{\rm BH}(M_1)} + \sqrt{S_{\rm BH}(M_2)}\right)^2 \,.
\label{TwoBlackHoles}
\end{equation}
The macroscopic quantum tunneling approach is actually another way of the derivation of the  Bekenstein-Hawking entropy of the black hole.

Maybe additivity could be restored if the black hole splitting were allowed at the classical level, i.e. for $\hbar\rightarrow 0$. But at the classical level, the tunneling process is forbidden. It is precisely because of the quantum processes that the ensemble of black holes has a special type of configuration space, where the entropy is not extensive. This non-extensivity is caused by quantum fluctuations, which determine the rare processes of macroscopic quantum tunneling between the black hole states. This is the analog of intermittency in the chaotic systems.\cite{Robledo2022,Elaskar2023}
Such processes require the generalization of the statistics with the corresponding non-extensive entropy. The equation (\ref{TwoBlackHoles}) fully determines the type of the statistics: it is described by the Tsallis-Cirto entropy with 
$\delta=2$, see Section \ref{TCentropy}.
Note, that this statistics is applicable to the black hole entropy as the entropy of the finite closed system.
It cannot be applied to such open systems as the de Sitter state, where the entropy is extensive.\cite{Volovik2024c,Volovik2024d}

 \section{Tsallis-Cirto  $\delta$-entropy for black hole}
\label{TCentropy}

The entropy in the black hole configurational space corresponds to the special class of the non-extensive entropies -- the Tsallis-Cirto $\delta$-entropy:\cite{TsallisCirto2013,Tsallis2020,Dabrowski2024}
\begin{equation}
S_\delta= \sum_{i=1}^n p_i \left(\ln \frac{1}{p_i}\right)^\delta\,,
\label{TsallisEntropy}
\end{equation}
where $ \sum_{i=1}^n p_i=1$.
For the Tsallis-Cirto entropy $S_\delta$, there is the following composition rule for two independent systems $A$ and $B$ with $n_{A+B}=n_A n_B$:
\begin{equation}
S^{1/\delta}_{\delta, A+B}= S^{1/\delta}_{\delta, A}+ S^{1/\delta}_{\delta, B}\,.
\label{Composition}
\end{equation}
 For $\delta=2$, this corresponds to the composition rule for entropies of black holes in Eq.(\ref{TwoBlackHoles}).

The full correspondence is reached if we assume that all the black-hole micro-states are equally probable with the probabilities
\begin{equation}
p_i=\frac{1}{n}= \exp\left(-\frac{M}{\sqrt{2}m_{\rm P}}\right)\,,
\label{EqualProbability}
\end{equation}
where  $m_{\rm P}$ is the reduced Planck mass,
\begin{equation}
m_{\rm P}=\frac{1}{\sqrt{8\pi G}} \,.
\label{PlanckMass}
\end{equation}
Then the equation (\ref{TsallisEntropy}) with $\delta=2$ gives the black hole entropy:
\begin{equation}
S_{\delta=2} =\left(\ln \frac{1}{p_i}\right)^2= \frac{M^2}{2m_{\rm P}^2}= 4\pi G M^2=S_{\rm BH}(M)\,.
\label{TsallisM}
\end{equation}

Note that the Tsallis-Cirto  $\delta =2$ entropy is the only example of the Renyi and Tsallis types of statistics, which exactly coincides with the Bekenstein-Hawking entropy. Any other alternative gives the wrong value of the Hawking temperature, as is discussed in Ref.\cite{Odintsov2021}, and the wrong Bekenstein-Hawking entropy.

 \section{Quantum of black hole mass}
 \label{QuantumMassSec}

Equation (\ref{EqualProbability}) can be interpreted in the following way. Let us introduce the "black-hole quantum" -- the black hole with the Planck-scale mass $M_0$:
\begin{equation}
M_0= \sqrt{2} m_{\rm P}= \frac{1}{\sqrt{4\pi G}}\,,
\label{massquantum2}
\end{equation}
so that the mass $M$ of the black hole is quantized:
\begin{equation}
 M=NM_0\,.
\label{Nquanta}
\end{equation}
Then from Eqs.(\ref{EqualProbability}) and (\ref{TsallisM}) one obtains that the probability $p_i$ and the Tsallis-Cirto $\delta=2$ entropy $S_{\delta=2}=S_{\rm BH}$ are expressed in terms of the number of quanta:
\begin{equation}
p_i=e^{-N} \,\,,\,\,  S_{\rm BH}=S_{\delta=2}=\ln^2 \frac{1}{p_i}=N^2\,.
\label{ProbabilityN}
\end{equation}
Thus the $\delta=2$ entropy provides the following composition rule for two ensembles with quanta $N_1$ and $N_2$:
\begin{equation}
S^{1/2}_{\delta=2} (N_1 + N_2)=N_1+N_2= S^{1/2}_{\delta=2}(N_1)+ S^{1/2}_{\delta=2}(N_2)\,.
\label{CompositionN}
\end{equation}

The probability $p_i$ in Eq.(\ref{ProbabilityN}) has the following physical meaning: its square is the rate of emission of a single black-hole quantum by the black hole with mass $M$:
\begin{equation}
p^2_i=w(M_0,M)=e^{-2N} \,.
\label{quantum}
\end{equation}
This can be seen from Eq.(\ref{HoleEmession1}) in the limit $N\gg 1$, where the entropy change after emission of the mass quantum $M_0$ is $N^2 - (N-1)^2 \rightarrow 2N$.

In this interpretation, the "elementary black hole" with the Planck-scale mass $M_0$ serves  as an "atom" of the micro-states in the black-hole configurational space. The number of the "atoms" forming the black hole with mass $M$ is $N=M/M_0$. This is valid for the black hole with arbitrary mass $M$, since the mass quantum $M_0$ does not depend on $M$. We shall see in Sec. \ref{RNsection} that this is valid also for the charged black holes.

The quantization of the black hole mass in Eq.(\ref{massquantum2}), $M=NM_0$, differs from the  Bekenstein quantization of entropy.\cite{Bekenstein1974,Mukhanov1986,Kastrup1997,Khriplovich2008,Dvali2011,Kiefer2020,Bagchi2024}
In the Bekenstein approach, the entropy is the adiabatic
invariant, which spectrum is equally spaced giving rise to the linear law, $S_{\rm BH}=a N$. Here $a$ is the dimensionless parameter, which value depends on the microscopic theory. The corresponding mass spectrum in this approach is $M\propto m_{\rm P}\sqrt{N}$.   In our approach, where only the black hole configurations are considered, the black hole entropy is quadratic,  $S_{\rm BH}=N^2$, while the mass spectrum is equidistant. The dimensionless parameter in this approach is the ratio between the mass quantum and the reduced Planck mass, $M_0/m_{\rm P}=\sqrt{2}$.

Note that in this approach, the ordinary matter is not considered. The role of matter is played by the ensemble of the "black hole atoms" -- the micro black holes with masses $M_0$ of Planck scale. As well as the ordinary particles, these elementary black holes may have spin and elementary electric charge, which allows to form the Reissner-Nordstr\"om and Kerr black holes.  Although the consideration in terms of elementary black holes looks rather artificial, it properly describes the black hole thermodynamics. The micro black holes with 
mass $M/m_{\rm P}=\frac{3}{2}$ were considered as the topological defects in the theory of the Multiple Point Principle.\cite{Nielsen2018,Nielsen2019}

Let us also mention the relation to the problems with the Hawking temperature $T_H$ discussed in Ref. \cite{Odintsov2022}.
The probability $p_i$ in Eq.(\ref{ProbabilityN}) can be represented in the following way:
\begin{equation}
p_i=e^{-N} =\exp\left(- \frac{M}{M_0}\right)=\exp\left(- \frac{M_0}{2T_H}\right) \,.
\label{ProbabilityN2}
\end{equation}
This looks as the probability of particle creation in the environment, which has the double Hawking temperature, $T=2T_H$. One may compare this with the doubling of the Hawking temperature considered by 't Hooft, \cite{Hooft2022,Hooft2023} and with the doubling of the Gibbons-Hawking temperature in the de Sitter environment.\cite{Volovik2024c,Volovik2024d,Maxfield2022} In both cases the doubling of temperature is related to the processes of co-tunneling, in which two processes of quantum tunneling occur simultaneously (coherently). In the 't Hooft scenario it is the simultaneous creation of particles and their quantum clones.  

In a similar way, the probability $w(M_0,M)=p^2_i$ in Eq.(\ref{quantum}) can be considered as the probability of the coherent creation of the "black hole atom" and its "clone" in the process of the co-tunneling:
\begin{equation}
w(M_0,M)=p^2_i=\exp\left(- \frac{M_0}{2T_H}\right) \exp\left(- \frac{M_0}{2T_H}\right)=\exp\left(- \frac{M_0}{T_H}\right)\,.
\label{cotunneling}
\end{equation}
This process looks as thermal with temperature $T_H$. However, in the 't Hooft approach, it can be considered as the combined process in which the separate processes (the process of creation of the black hole atom and the process of creation of its clone) are determined by the temperature $T=2T_H$.

\section{Application to Reissner-Nordstr\"om black hole}
\label{RNsection}

Let us show that this approach is applicable also to the electrically charged Reissner-Nordstr\"om (RN) black holes. In the RN black hole, the positions of its two horizons, $r_+$ and $r_-$, are expressed in terms the mass $M$ of the black hole and its charge $q$:   
 \begin{equation}
r_+r_-= \alpha q^2G\,\,, \,\, r_+ + r_-= 2MG\,.
\label{r+r-}
\end{equation}
Here, the charge $q$ is in units of the electric charge of electron and $\alpha$ is the fine structure constant.

In Ref.\cite{Volovik2021a} it was shown that the entropy of the RN black hole does not depend on charge $q$ and is only determined by its mass: 
 \begin{equation}
S_{\rm RN}(M,q,\alpha)=S_{\rm RN}(M,q,\alpha=0)=S_{\rm BH}(M)= 4\pi GM^2 \,.
\label{entropyRNtilde}
\end{equation}
This follows in particular from the observation that by varying the fine structure constant $\alpha$ the RN black hole can be adiabatically transformed to the Schwarzschild black hole with the same mass $M$ and the same charge $q$.  Indeed, for $\alpha=0$, the presence of the quantized electric charge $q$ does not influence the black hole and its entropy.

On the other hand, the entropy in Eq.(\ref{entropyRNtilde}) can be expressed in terms of the entropies $S_-=\pi r_-^2/G$ and $S_+ =\pi r_+^2/G$ of the inner and outer horizons:
\begin{equation}
S_{\rm RN}= \left( \sqrt{S_+} + \sqrt{S_-}\right)^2 \,.
\label{entropyRNtilde2}
\end{equation}
This also agrees with the non-extensive Tsallis-Cirto entropy with $\delta=2$,
which supports the special properties of the black hole configurational space.

The Hawking radiation from the RN black hole is characterized by the modified Hawking temperature, which also does not depend on the charge $q$:
 \begin{equation}
T_{\rm H} = \frac{1}{4\pi (r_+ +r_-)}= \frac{1}{8\pi G M} \,.
\label{HawkingTtilde1}
\end{equation}
This modified Hawking temperature agrees with the thermodynamic equation:
 \begin{equation}
dS_{\rm RN} =  d(4\pi GM^2)=\frac{dM}{T_{\rm H}} \,.
\label{ThermodynamicsTtilde}
\end{equation}

One may extend the consideration to the rotating Kerr black hole. It has been suggested that its entropy is also determined by the mass $M$ and does not depend on the black hole angular momentum $J$.\cite{Okamoto1992} In this case the contribution of the angular momentum can be nullified after adiabatic transformation by varying the Planck constant $\hbar \rightarrow \infty$.\cite{Volovik2021a} The final state of the Kerr black hole has same mass $M$ and the same angular momentum  $J$, but its entropy becomes the same as the  Schwarzschild black hole with $J=0$. 
If it is so, then the mass and entropy of the RN and Kerr black holes obey the same quantization rules as the Schwarzschild black hole, $M=NM_0$ and $S=N^2$, with the same value of the mass quantum $M_0=\sqrt{2}m_{\rm P}$. This differs from the other quantization schemes. The approach in which the horizon area is considered as adiabatic invariant\cite{Barvinsky2001,Barvinsky2002,Vagenas2011,Volovik2022} gives the linear quantization rule, such as $S=2\pi N$. The quantization inspired by the string theory\cite{Gibbons2011,Visser2012} gives the square-root rule,
$S =2\pi(\sqrt{N_1} \pm \sqrt{N_2})$. 

Another possible extension is the thermodynamics of white holes. Their entropy is negative as follows from the same macroscopic quantum tunneling approach.\cite{Volovik2022} One possibility is to introduce in addition to the elementary black holes also the white-hole "atoms" with the same value $M_0$ of the elementary mass quantum. Then the entropy of the white hole with mass $M=M_0N$ is $S_{\rm WH}(N)=-N^2$, and the transition probability from the black hole to the white hole of the same mass\cite{Rovelli2019,Volovik2022} is $w\propto \exp(S_{\rm WH}-S_{\rm BH})= \exp(-2S_{\rm BH})=\exp(-2N^2)$.

\section{Conclusion}
\label{ConclusionSec}

The quantum tunneling processes related to the black hole determine the black hole thermodynamics. The Hawking temperature is determined by the quantum tunneling processes of radiation of particles from the black hole. The Bekenstein-Hawking entropy of the black hole is obtained by consideration of the macroscopic quantum tunneling processes of radiation of small black holes from the large one.
These tunneling processes also determine the composition rule for the black hole entropy. This entropy is not extensive and obeys the composition rule in Eq.(\ref{TwoBlackHoles}), which corresponds to the Tsallis-Cirto entropy with $\delta=2$ in Eq.(\ref{Composition}).

The macroscopic quantum tunneling processes of splitting of black hole to the smaller parts suggests that the black hole  can be considered as an ensemble of the elementary black holes with the Planck-scale mass $M_0=1/\sqrt{4\pi G}$. The black hole mass is $M=NM_0$, where $N$  is the number of these micro black holes, which play the role of the black hole "atoms". The  Bekenstein–Hawking entropy of the black hole with $N$ atoms, is $S_{\rm BH}(N)=N^2$, with the Tsallis-Cirto $\delta=2$ composition law in Eq.(\ref{CompositionN}). 

{\bf Acknowledgements}. I thank Sergey Odintsov for discussions.

\end{document}